\documentclass[10pt]{article}
\usepackage{amsfonts}
\usepackage{multicol}
\usepackage{amsmath}
\usepackage{amssymb}
\usepackage{mathptmx}
\usepackage{cite}
\usepackage{slashed}
\usepackage{epsfig}
\usepackage{graphicx}
\usepackage{array}
\usepackage{multirow}
\usepackage{color}
\usepackage[colorlinks=true,linkcolor=red,citecolor=blue]{hyperref}
\usepackage{url}
\begin{document}
\newcommand{\beq}{\begin{eqnarray}}
\newcommand{\eeq}{\end{eqnarray}}
\newcommand{\non}{\nonumber\\ }
\newcommand{\acp}{ {\cal A}_{CP} }
\newcommand{\psl}{ p \hspace{-1.8truemm}/ }
\newcommand{\nsl}{ n \hspace{-2.2truemm}/ }
\newcommand{\vsl}{ v \hspace{-2.2truemm}/ }
\newcommand{\epsl}{\epsilon \hspace{-1.8truemm}/\,  }
\newcommand{\tf}{\textbf}
\title{ Recent Anomalies in B Physics}
\author{Ying Li$^1$\footnote{liying@ytu.edu.cn},\,\,\, Cai-Dian L\"u$^2$\\
{\small \it 1. Department of Physics, Yantai University, Yantai 264005,China}\\
{\small \it 2. Institute of High Energy Physics, Beijing 100049, China}\\
}
\maketitle
\begin{abstract}
$B$ physics plays important roles in searching for the new physics (NP) beyond the standard model (SM). Recently, some deviations between experimental data and SM predictions were reported, namely  $R(D^{(*)})$, $P_5^\prime$ and $R_{K^{(*)}}$ anomalies. If these anomalies were further confirmed in future,  they would be unambiguous hints  of NP.  Theoretically, in order to explain these anomalies, a large number of models have been proposed, such as models including  leptoquark or $Z^\prime$.  However, these new particles have not been discovered directly in LHC. Moreover, the models should pass the examination of $B_s\to \mu^+\mu^-$ and $B_s^0-\bar B_s^0$ mixing. In future, the analysis of data taken during the ongoing Run 2 of the LHC and the  forthcoming Belle-II will present new insight both into the observables of interest and into new strategies to control uncertainties. Theoretically, the existed  models should be further tested; and more NP models are welcomed to explain these anomalies simultaneously without affecting other measurements consistent with SM.
\end{abstract}
\par  It is widely believed that Standard model (SM) is not a final theory and should be an effective one of new theory at higher scale, because SM has a large number of parameters that have to be input from experiment, it does not provide any candidate for dark matter, dark energy and the resource of  the matter-antimatter asymmetry of the universe, and does not incorporate gravity.  For this reason, the major task of the particle physics switched to the search for the new physics (NP) beyond SM by probing the signals directly and testing SM precisely, after the discovery of Higgs in 2012. However, it is very depressing for us that there is still no strong evidence of  NP on the two experiments (ATLAS and CMS) at the CERN Large Hadron Collider (LHC) up to date.
In contrast, the LHCb experiment could study the properties of particles containing heavy quarks with high precision, with the large cross section for $B$ meson production in $pp$ collisions . Prior to the startup of LHCb, two $B$ factory experiments, BaBar and Belle, had dominated the landscape of the heavy flavour physics. The  measurements of $CP$ violation in $B$ sector confirmed SM expectations and led to Nobel prizes for Kobayashi and Maskawa in 2008. Furthermore, many notable results from these experiments, which not only tested SM, such as the CKM triangle,  but placed stringent constraints on NP, such as,  measurements of $B$ mesons decays to final states containing $\tau$ leptons, i.e., purely leptonic $B^+ \to \tau^+\nu_\tau$, the mass differences of neutral mesons and the semileptonic decay $B\to K^{(*)}\ell^+\ell^- $. Recently, LHCb and Belle experiments  reported a few anomalous results in $B$ sector, which cannot be accommodate in SM and might be  hints of NP. In this short review, we will introduce these anomalies in detail.

\begin{figure*}[!htb]
\centering
\includegraphics[width=0.5\textwidth]{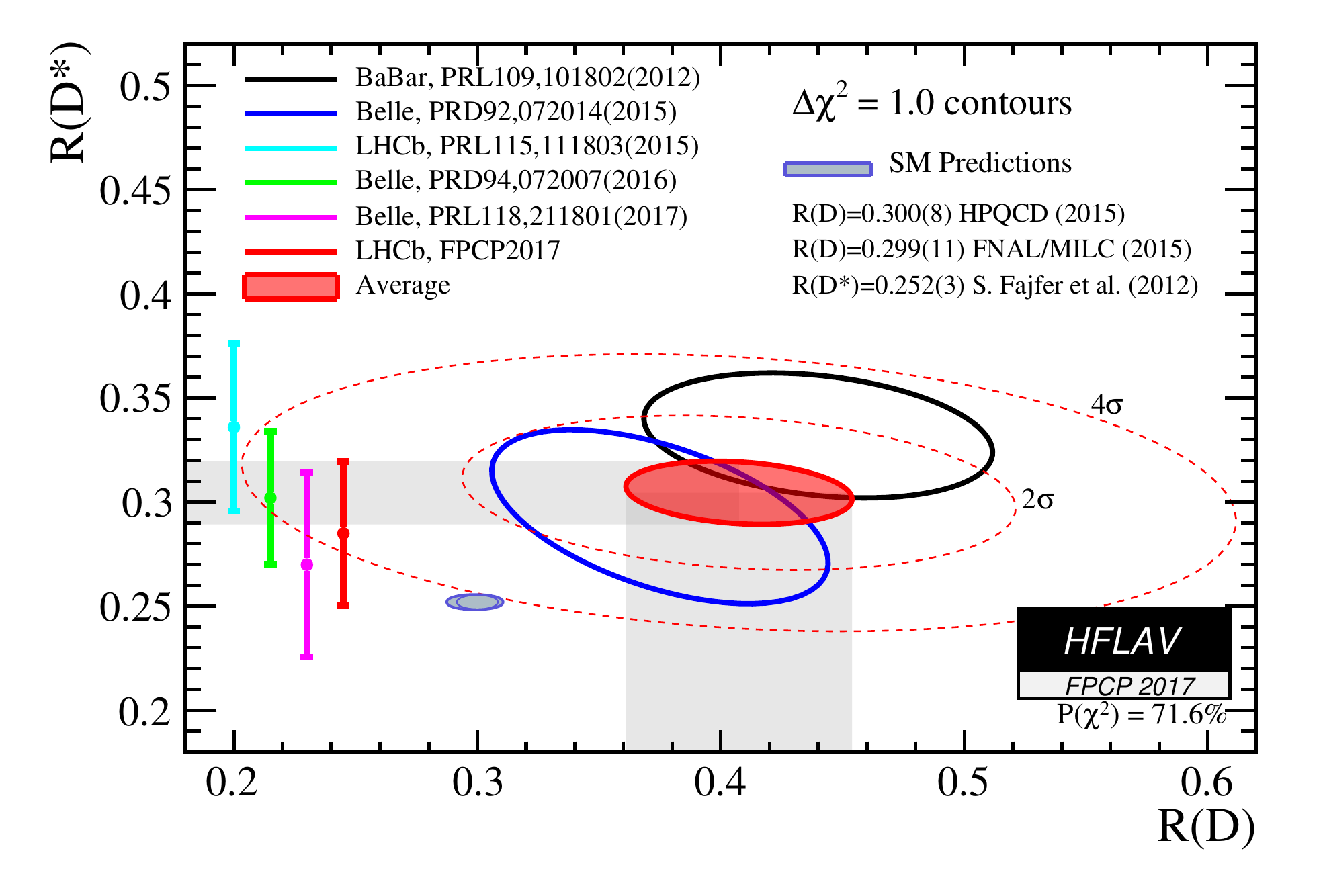}
\caption{Measurements of $R(D^{(*)})$~\cite{Lees:2012xj,Lees:2013udz, Huschle:2015rga,
Aaij:2015yra, Sato:2016svk, Abdesselam:2016xqt,Hirose:2016wfn }, their averages~\cite{HFAG}, the
SM predictions~\cite{Bernlochner:2017jka, Lattice:2015rga, Na:2015kha,
Aoki:2016frl,Fajfer:2012vx}}.\label{fig:RDdata}
\end{figure*}

In SM, the semileptonic decays of $B$ mesons induced by $b\to c \ell\nu$ can be well understood. Specially, these decays involving $\tau$ lepton can be used to prob the intermediate charged Higgs boson or other non-SM particles, in that the higher mass of $\tau$ lepton leads to additional amplitudes. In order to reduce the hadronic uncertainties and some of the experimental systematic errors, the typical parameters $R(D^{(*)})$ are defined as
\begin{eqnarray}
R(D^{(*)}) =\frac{\Gamma(B\to D^{(*)}\tau\bar\nu)}{\Gamma(B\to D^{(*)}\ell \bar\nu)}
\end{eqnarray}
where $\ell=\mu,e$. The current world averaged measurements \cite{HFAG} and SM expectations \cite{Aoki:2016frl, Fajfer:2012vx} are given as
\begin{eqnarray}
R(D)&=&\left\{
               \begin{array}{ll}
                 0.407 \pm 0.039\pm 0.024, & \hbox{Exp.\cite{HFAG};} \\
                 0.300 \pm 0.008, & \hbox{SM\cite{Aoki:2016frl}};
               \end{array}
             \right.
\\
R(D^{*})&=&\left\{
               \begin{array}{ll}
                 0.304 \pm 0.013\pm 0.007, & \hbox{Exp.\cite{HFAG};} \\
                 0.257 \pm 0.005, & \hbox{SM\cite{Fajfer:2012vx}}.
               \end{array}
             \right.
\end{eqnarray}
The experimental results and SM predictions are also showed in Figure.~\ref{fig:RDdata}. It is apparent that these independent measurements show good consistency with one another. Moreover, the theoretical calculations are also on solid footing, as heavy quark symmetry suppresses  the uncertainties from the hadronic physics needed for SM prediction. From the figure, it is also seen that $R(D)$ and $R(D^*)$ exceed SM predictions by $2.3\sigma$ and $3.4\sigma$,  respectively.  Considering the $R(D)-R(D^*)$ correlation of $-0.20$, the resulting combined $\chi^2$ is 20.60 for 2 degree of freedom, corresponding to a $p$-value of $4.13 \times 10^{-5}$ \cite{HFAG}. The difference with SM predictions corresponds to about $4.1\sigma$,  which is the most largest discrepancy from SM in particle physics.

In the theoretical view, it is very strange for us to find such large deviations from SM in these processes that occur at tree level. If these measurements can be further confirmed by future experiments, which would imply that the scale of the NP may be at or below the TeV scale. Currently, at ATLAS and CMS, many NP models have been excluded by LHC Run 1 bounds already, and more will soon be constrained by the Run 2 data. To accommodate the current experimental values, many NP scenarios with leptoquark, $W^\prime$, vector quark, charged scalars, or lepton mixing have been proposed. Model-independent analysis indicates that the vector type particles are preferred, compared to scalars. Moreover, in order to agree with other existed experimental data and avoid the current constraints from direct production at LHC experiment, leptoquarks that mainly couples to the third generation of fermions are favored. In addition, there are some viable scenarios in which $B\to D^{(*)}\tau\bar\nu$ are SM-like, but $B\to D^{(*)} \ell \bar\nu$ are suppressed by interference between NP and SM.

In order to clarify this anomaly, more further decay processes should be measured, such as $B_c \to J/\psi \tau\bar\nu$ vs. $B_c \to J/\psi \mu\bar\nu$, $\Lambda_b \to \Lambda_c \tau\bar\nu$ vs. $\Lambda_b \to \Lambda_c \mu\bar\nu$, as well as the $B\to D^{**}\tau\bar\nu$ rates.  Very recently, LHCb released their first measurement of $Br(B_c^+\to J/\psi \mu^+\nu_\mu)/Br(B_c^+\,\to J/\psi \tau^+\nu_\tau)$ \cite{Aaij:2017tyk}. The ratio of the branching fractions is measured to be $R(J/\psi)=0.71 \pm 0.17(stat) \pm 0.18(syst)$, which lies within $2\sigma$ deviations above the range of existing predictions in SM. Other decay modes will be studied in the  LHCb and the forthcoming Belle-II with fully differential theory predictions.  Once a deviation from SM is established, it would motivate us to measure all other possible semileptonic $B$ decays with $\tau$ lepton, both in $b\to c$ and $b\to u$ transitions.

Flavour-changing neutral-current (FCNC), such as $b\to s \ell^+\ell^-$,  also constitute sensitive probes of NP beyond SM, since it  is forbidden at tree-level and can only occur at loop order in SM. The new heavy particles might appear in competing diagrams and presents sizable contribution. In particular, the effective Hamiltonian of the $b\to s \ell^+\ell^-$ transition  allow us to separate short and long distances, and the short distances are encoded in the Wilson coefficients of the relevant operators and long distance are in the matrix elements of these operators, which  are given as:
\begin{eqnarray}
&&O_7 = \frac{e}{16 \pi^2}m_b\,  \bar s\sigma^{\mu\nu}(1+\gamma_5)F_{\mu\nu}\,b  \quad {\rm [real \,\, or \,\, soft \,\, photon]}\\
&&O _9=\frac{e^2}{16 \pi^2}\bar{s}\gamma_\mu(1-\gamma_5)b\  \bar\ell\gamma^\mu\ell
 \quad  {\rm [ Z/hard \,\,  \gamma \ldots]}\\
&&O_{10}=\frac{e^2}{16 \pi^2}\bar{s}\gamma_\mu(1-\gamma_5)b\  \bar\ell\gamma^\mu\gamma_5\ell
\quad  {\rm [  Z ]}
\end{eqnarray}
NP either induces a change in the Wilson coefficients by adding a new contribution or generate new operators (chirally flipped, scalar or pseudoscalar or tensor operators). 

\begin{figure*}[!htb]
\centering
\includegraphics[width=0.45\textwidth]{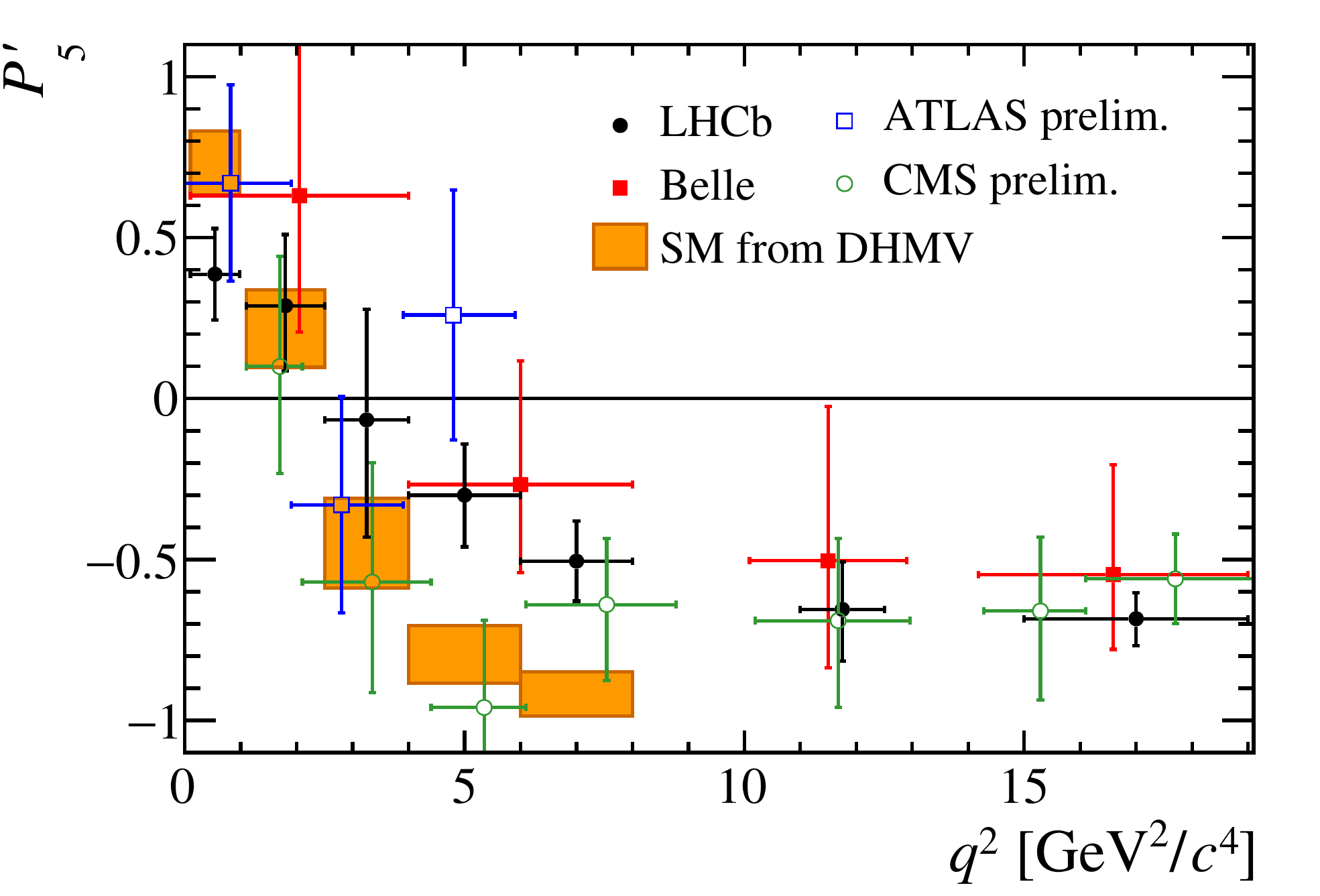}\hfill
\caption{ The theoretical predictions \cite{Descotes-Genon:2014uoa} and experimental results \cite{Aaij:2015oid, Abdesselam:2016llu, ATLAS-CONF-2017-023,CMS-PAS-BPH-15-008} of  $P_5'$  in $B\to K^{*} \mu^+\mu^-$ decay. }\label{fig:2}
\end{figure*}

As a typical process induced by $b\to s\ell^+\ell^-$, $B\to K^{*} \ell^+\ell^-$ decay has been paid much attention.  Because the final state $K^{*}$ is a vector meson,  it is of interest for us to investigate  the  angular  distribution in detail for searching for observables that are sensitive to the NP contribution. The resulting angular distribution can be parameterized in terms of eight angular observables.  Among them, observable $P_5^\prime$ has usually been viewed as a good probe of NP due to few hadronic uncertainties, in term of the analysis based on the  soft-collinear effective theory.  However, recent measurements showed that a tension exists between experimental results of the  $P_5'$ and their corresponding SM prediction in the region $\rm [4 , 8]Gev^{2}$, as illustrated in Fig.~\ref{fig:2}. Note that the data from ATLAS, Belle and LHCb are above SM predictions remarkably. The CMS result is more consistent with SM. In SM,  the contribution from the high power corrections, such as charm loop and soft gluon interactions,   can only alleviate the tension and cannot explain this anomaly. 

Apart from the $P_5^\prime$ anomaly, another ones are the violations of the lepton universality in $B$ sector, which are denote by \begin{eqnarray}
R_{K^{(*)}}=\frac{\Gamma(B\to K^{(*)}\mu^+\mu^-)}{\Gamma(B\to K^{(*)}e^+e^-)}.
\end{eqnarray}
In SM, they are protected from hadronic uncertainties and can be very accurately predicted, because the mass difference between electron and muon is negligible in ${ O}(m_b)$ scale. In 2014, the LHCb collaboration reported their first measurement ~\cite{Aaij:2014ora}
\begin{eqnarray}
R_K=0.745^{+0.090}_{-0.074}(stat)\pm0.036(syst), q^2 \in [1,6]~\text{GeV}^2,
\end{eqnarray}
which deviated from SM predictions $R_K=1.0004(8)$~\cite{Bordone:2016gaq} with a significance of $2.6\sigma$, $q^2$ standing for the dimuon invariant mass squared. In 2017, LHCb released another measurements in two bins of $q^2$~\cite{Aaij:2017vbb}, 
\begin{eqnarray}
&&R_{{K^*}}=0.660^{+0.110}_{-0.070}(stat)\pm0.024(syst),~~q^2 \in [0.045,~1.1]~\text{GeV}^2\\
&&R_{{K^*}}=0.685^{+0.113}_{-0.069}(stat)\pm0.047(syst),~~q^2 \in [1.1,~6]~\text{GeV}^2
\end{eqnarray}
which show again deficits with respect to SM predictions \cite{Geng:2017svp}
\begin{eqnarray}
&&R_{{K^*}}^{\rm SM}=0.920(7),~~q^2 \in [0.045,~1.1]~\text{GeV}^2,\\
&&R_{{K^*}}^{\rm SM}=0.996(2),~~ q^2 \in [1.1,~6]~\text{GeV}^2,
\end{eqnarray}
with a significance of $2.3\sigma$ and $2.4\sigma$, respectively.   

Furthermore, large discrepancy was reported in $B_s^0 \to \phi \mu^+\mu^-$ decay, which is also induced by the $b\to s\ell^+\ell^-$ transition,  and the data are more than $3\sigma$ from SM predictions in the $q^2\in[1,~6]$ GeV$^2$ \cite{Aaij:2015esa}. Unfortunately, in SM, we have not found extra contributions that could account for above large deviations till now.

Although none of these anomalies is yet significant on its own, the situation is quite intriguing because all affected decays are induced by $b\to s\ell^+\ell^-$ and thus probe the same NP. The model-independence global analysis results showed that additional scalar or pseudoscalar couplings can not address the tension as their contributions are suppressed by small lepton masses. Therefore, as for the NP, the natural models to introduce another gauge  boson, namely $Z^\prime$, with family non-universal couplings. In the literature, the most popular one is based on gauging $L_\tau - L_\mu$ lepton number \cite{Altmannshofer:2014cfa}, in which the vanishing coupling of the $Z^\prime$ to electrons can explain  $R_{K^{*}}$.  An alternative scenario is to introduce the scalar or vector leptoquarks that couple leptons to quarks via new vertices. Different leptoquark models can be classified according to the spin of the leptoquarks and their quantum numbers with respect to SM gauge groups. Because the conservation of lepton flavour is not respected, leptoquark models could explain the $R(D^{(*)})$ and $R_{K^{(*)}}$ simultaneously. In the past years, various representations of leptoquarks and $Z^\prime$ have been studied  extensively in literatures. 

It should be noted that the new particle generating $b\to s\ell^+\ell^-$  also contributes to $B_s^0 \to \ell^+\ell^-$ decays. Very recently, LHCb measured the branching fraction of  $B_s^0 \to \mu^+\mu^-$ \cite{Aaij:2017vad}. They reported a branching fraction for this process of $(3.0\pm0.6\pm0.3)\times10^{-9}$ , the rarest observed heavy flavour branching fraction to date, which agrees with predictions of  SM. This result will constrain the parameters of the leptoquark models, such as the mass of leptoquark and the couplings between quarks and leptons stringently. Furthermore, the new particles also contribute to the neutral meson mixings, such as the mass difference $\Delta m_{B_s}$. For instance, in models with neutral gauge boson $Z^\prime$, both $b\to s\ell^+\ell^-$ and $B_s^0- \overline B_s^0$ mixing could be  generated by tree-level exchange of the $Z^\prime$ boson.  The precise measurements of $\Delta m_{B_s}$ could constrain the couplings of $Z^\prime-b-s$, as well as the mass of $Z^\prime$. As for the direct searches at LHC, we have not found any signals of NP till now, as aforementioned.

In summary,  several deviations from SM expectations have been reported in $B$ physics in recent years, including the violation of lepton flavour universality in $b \to c \ell \nu$ transitions with $\ell=\tau$ vs. $\mu$ or $e$,  the  violation of lepton flavour universality in $b \to s \ell ^+\ell^-$ transitions with $\ell = e$ vs $\mu$,  as well as $P_5^\prime$ anomaly in $B\to K^{*}\ell^+\ell^- $ decay. If these results were furthered confirmed, they would constitute unambiguous evidence of NP beyond SM, and could be regarded as the hints of NP. Theoretically, possible explanations involving particles beyond SM exist in the form of leptoquarks or $Z^ \prime$ bosons.  In addition, the decays $B_s^0 \to \ell^+\ell^-$, the mass differences of neutral mesons and direct searches pose tight constraints on some of these models.  In future, the analysis of data taken during the ongoing Run 2 of the LHC and the  forthcoming Belle-II will present new insight both into the observables of interest and into new strategies to control uncertainties. For the Run 2 of LHC, we shall continue to measure other lepton flavor violation decays and more related observables; while for Belle-II with high luminosity, the uncertainties are hoped to be reduced. Furthermore, another important work in experiments is to search for the predicted particles directly on the Run 2 of the LHC, so as to test the models aforementioned. Of course, in the theoretical side, on the one hand we should study the existed models comprehensively and present more observables for testing on the experiments; on the other hand, more NP models are welcomed to explain these anomalies simultaneously without affecting other measurements consistent with SM.
\section*{Acknowledgment}
We  thus here acknowledge all researchers who contribute to these studies, and we cannot cite all references related to this work due to the limited article space. This work was supported in part by the National Science Foundation of China under the Grant Nos.~11575151,11235005, and by the Natural Science Foundation of Shandong province (ZR2016JL001).


\end{document}